\documentclass[runningheads]{llncs}
\usepackage{todonotes}
\usepackage{cite}

\makeatletter
\newcommand{\printfnsymbol}[1]{%
  \textsuperscript{\@fnsymbol{#1}}%
}
\makeatother
\usepackage{amsmath}
\usepackage{booktabs}
\usepackage{tabularx}
\usepackage{amssymb} 
\usepackage{multirow}
\usepackage[pagebackref,breaklinks,colorlinks]{hyperref}
\usepackage[normalem]{ulem} 
\usepackage{tablefootnote}
\usepackage{orcidlink}

\usepackage{graphicx}
\usepackage{caption}
\usepackage{subcaption}
\begin{document}
%
\title{Longitudinal Segmentation of MS Lesions via Temporal Difference Weighting}
\titlerunning{Longitudinal Segmentation of MS Lesions via Temporal Difference Weighting}
%
\author{Maximilian R. Rokuss\inst{1,2}\orcidlink{0009-0004-4560-0760}\thanks{Contributed equally. Each co-first author may list themselves as lead author on their CV.} \and
Yannick Kirchhoff\inst{1,2,3}\orcidlink{0000-0001-8124-8435}\printfnsymbol{1} \and
Saikat Roy\inst{1,2}\orcidlink{0000-0002-0809-6524} \and
Balint Kovacs\inst{1,4}\orcidlink{0000-0002-1191-0646} \and
Constantin Ulrich\inst{1,4}\orcidlink{0000-0003-3002-8170} \and
Tassilo Wald\inst{1,2,5}\orcidlink{0009-0007-5222-2683} \and
Maximilian Zenk\inst{1,4}\orcidlink{0000-0002-8933-5995} \and
Stefan Denner\inst{1,4}\orcidlink{0009-0008-2302-743X} \and
Fabian Isensee\inst{1,5}\orcidlink{0000-0002-3519-5886} \and
Philipp Vollmuth\inst{1,6,7}\orcidlink{0000-0002-6224-0064} \and
Jens Kleesiek\inst{8,9}\orcidlink{0000-0001-8686-0682} \and
Klaus Maier-Hein\inst{1,10}\orcidlink{0000-0002-6626-2463}
}


%
\authorrunning{M. Rokuss, Y. Kirchhoff et al.}
%
\institute{German Cancer Research Center (DKFZ), Heidelberg, Division of Medical Image Computing, Germany
\and
Faculty of Mathematics and Computer Science, Heidelberg University, Germany
\and
HIDSS4Health - Helmholtz Information and Data Science School for Health, Karlsruhe/Heidelberg, Germany \and
Medical Faculty Heidelberg, Heidelberg University, Heidelberg, Germany
\and
Helmholtz Imaging, German Cancer Research Center, Heidelberg, Germany
\and
Department of Neuroradiology, Heidelberg University Hospital, Heidelberg, Germany
\and
Division for Computational Neuroimaging, Department of Neuroradiology, Heidelberg University Hospital, Heidelberg, Germany
\and
Institute for Artificial Intelligence in Medicine (IKIM), University Hospital Essen, Essen, Germany
\and
Cancer Research Center Cologne Essen (CCCE), West German Cancer Center Essen, University Hospital Essen, Essen, Germany
\and
Pattern Analysis and Learning Group, Department of Radiation Oncology, Heidelberg University Hospital\\
\email{\{maximilian.rokuss,yannick.kirchhoff\}@dkfz-heidelberg.de}
}
\maketitle              
\begin{abstract}

Accurate segmentation of Multiple Sclerosis (MS) lesions in longitudinal MRI scans is crucial for monitoring disease progression and treatment efficacy. Although changes across time are taken into account when assessing images in clinical practice, most existing deep learning methods treat scans from different timepoints separately. Among studies utilizing longitudinal images, a simple channel-wise concatenation is the primary albeit suboptimal method employed to integrate timepoints. We introduce a novel approach that explicitly incorporates temporal differences between baseline and follow-up scans through a unique architectural inductive bias called Difference Weighting Block. It merges features from two timepoints, emphasizing changes between scans. We achieve superior scores in lesion segmentation (Dice Score, Hausdorff distance) as well as lesion detection (lesion-level F\textsubscript{1} score) as compared to state-of-the-art longitudinal and single timepoint models across two datasets. Our code is made publicly available at: \url{www.github.com/MIC-DKFZ/Longitudinal-Difference-Weighting}.


\keywords{Medical Image Segmentation \and Longitudinal Imaging \and Multiple Sclerosis}
\end{abstract}

\section{Introduction}

Multiple sclerosis (MS) is an inflammatory disease in the central nervous system (CNS), characterized by the accumulation of demyelinating white-matter lesions in the brain and spinal cord, especially affecting young adults~\cite{multiple_sclerosis}. Magnetic resonance imaging (MRI) has been proven as a valuable diagnostic tool for identifying and monitoring MS. In clinical practice, patients typically undergo routine screenings that involve multiple MRI scans throughout their treatment. For instance, official MS guidelines advocate for regular screenings at intervals ranging from 3 to 12 months \cite{MS_FollowUp_Guidelines}. Monitoring the progress of MS lesions is crucial for evaluating the effectiveness of anti-inflammatory disease-modifying drugs~\cite{msseg2}. However, analyzing the lesion load of several MS scans at different timepoints is a heavy burden for clinicians, adding to their workload~\cite{coactseg}. Thus, automating lesion segmentation is critical for clinical computer-aided diagnosis (CAD) systems, facilitating precise quantification for evaluating treatment response~\cite{ms_decision_model}.

There has been a multitude of approaches proposed for the automated segmentation of MS lesions in brain MRI sequences, with deep learning being widely utilized~\cite{review_dl_ms_lesion_seg, review_ms_lesion_seg}. However, common methods for white-matter lesion segmentation process images from various timepoints independently~\cite{icobrain,Brugnara2020}, deviating from the clinical practice of comprehensively assessing patients' progression over time. Current state-of-the-art biomedical image segmentation pipelines like nnUNet~\cite{Isensee2021} or SwinUNETR~\cite{swinunetr} are cross-sectional methods, meaning they only use information from single timepoints, even when longitudinal (multiple timepoint) data is available. So far, only a few works proposed using the information of prior scans for the segmentation task~\cite{review_ms_longitudinal_segmentation}. Early work from Birenbaum et al.~\cite{Birenbaum} passes 2D slices from multiple timepoints through a single view CNN, which are then concatenated and used to classify the center pixel. Denner et al.~\cite{denner2020spatiotemporal} proposed a multitask network simultaneously trained on registration and segmentation using a 2.5D approach. Szeskin et al.\cite{simunet} employed a multi-channel 3D recurrent residual U-Net model (R2UNet~\cite{R2UNet}) trained on pairs of registered scans, benefiting from the inclusion of additional time steps as inputs. Wu et al.~\cite{coactseg} combine heterogeneous dataset annotations using a model that outputs the new-lesion and all-lesion mask given a baseline and follow-up scan. Notably, the non-deep learning-based neuroimage analysis tool \textit{FreeSurfer}~\cite{free_surfer} was recently extended with a longitudinal whole-brain and white-matter lesion segmentation method, solving an optimization problem across all scans of a given patient~\cite{free_surfer_longi}. 

Nevertheless, none of the above methods impose an explicit architectural inductive bias to guide the network to use the additional timepoint. While multiple timepoints can be trivially incorporated into standard UNet-like architectures by channel-wise concatenation~\cite{simunet,coactseg,denner2020spatiotemporal}, a potential concern is that the network might primarily emphasize one particular scan similar to a single timepoint model. Previous methods lack thorough evaluation against strong benchmarks\cite{denner2020spatiotemporal,simunet}, often comparing only to their own method with single timepoint input or basic U-Net architectures, without considering state-of-the-art single timepoint methods like nnUNet~\cite{Isensee2021} or SwinUNETR~\cite{swinunetr}. Additionally, a recent review revealed that most longitudinal lesion segmentation models lack publicly available datasets or code, negatively impacting reproducibility~\cite{review_ms_longitudinal_segmentation}.

Addressing the aforementioned issues, we propose a method for longitudinal MS lesion segmentation, inherently adhering to clinical practice of assessing an image by comparison, i.e. the difference, to the previous scan. Our method introduces explicit inductive bias to leverage the information surplus from the additional baseline scan. Specifically, our model encodes both baseline and follow-up images into a latent space where a novel \textit{Difference Weighting Block} produces a temporally-informed combined representation which the decoder uses to generate the segmentation mask. Extensive experiments on two public MS datasets from different sites demonstrate that our proposed model outperforms all state-of-the-art single timepoint baselines as well as all publicly available longitudinal MS lesion segmentation methods. It achieves superior performance for both the pixel-based Dice score and the clinically more relevant lesion-based F\textsubscript{1} score, thus improving the lesion detection rate.\\

Overall, our contributions encompass these key aspects:

\begin{itemize}
        \item \textbf{Pitfalls of existing Longitudinal Methods:} We demonstrate that state-of-the-art single timepoint baselines surpass existing longitudinal methods. However, even naively combining longitudinal data with strong single timepoint models yields superior performance.
        \item \textbf{Difference Weighting Block:} Inspired by clinical practice, we incorporate explicit architectural bias to fully leverage the benefits of longitudinal information. We introduce a novel component for merging features from different timepoints in latent space, significantly enhancing generated segmentations.
        \item \textbf{Cross-Dataset Generalization of Longitudinal Benefits:} Our method demonstrates transferability to unseen datasets, indicating that the advantages of longitudinal information can be effectively applied to real world scenarios.
\end{itemize}

\section{Method}

\subsection{Integrating Prior Images for Enhanced Segmentation}

Similar to a single timepoint method, our model $F_{\theta}$ is designed to predict a segmentation mask $\hat{Y}_c$ for a \textit{current} scan $X_c$. However, it additionally incorporates information from a \textit{prior} scan $X_p$ by co-learning both representations. The method thus leverages the change between $X_p$ and $X_c$ as depicted in Eq. \ref{eq:model}. For instance, the \textit{prior} image can correspond to the baseline scan, and the \textit{current} image represents a subsequent follow-up examination. For predicting the baseline scan we use the subsequent follow-up examination as additional information.

\begin{equation}
    F_\theta: \mathbb{R}^3 \to \mathbb{R}^3,\, \hat{Y}_c = F_{\theta}(X_c, X_p) \text{ where } X_c \neq X_p
    \label{eq:model}
\end{equation}\\

\begin{figure}[t]
    \includegraphics[width=\textwidth]{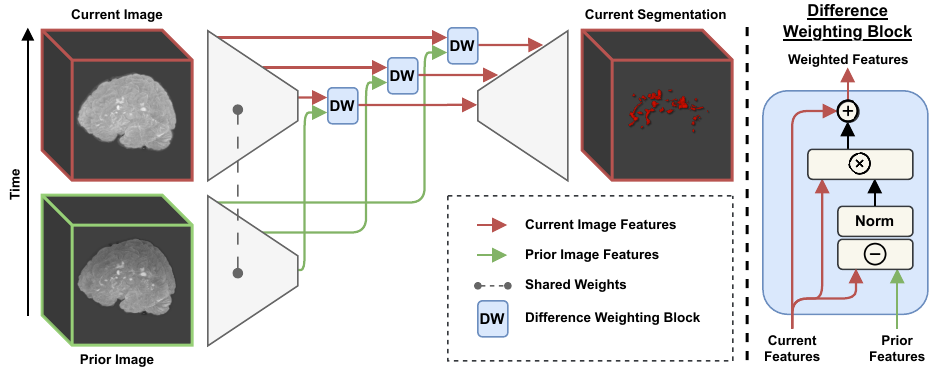}
    \caption{\textit{Left:} Pipeline of our proposed model. First, the prior image (baseline) and current image (follow-up) are both encoded by a shared encoder. Subsequently, the Difference Weighting Block is applied on all skip connections to merge these encoded representations, emphasizing dissimilarities between the two images. This temporally enhanced representation is then fed to the decoder to produce the segmentation of the current image. \textit{Right:} Architecture of the Difference Weighting Block. Firstly, the difference between the features extracted from the current and prior images is computed, followed by an Instance Norm. This result is then multiplied with the current image features and finally a residual connection is added.}
    \label{fig:pipeline}
\end{figure}

\noindent The aim of a longitudinal approach is to improve the segmentation by utilizing information from multiple timepoints. Instead of the early fusion of longitudinal scans in the form of channel-wise concatenation, we propose a later fusion of features from the two timepoints. We achieve this by passing both the baseline and follow-up images through a shared encoder. Then, we merge the resulting features using a novel \textit{Difference Weighting Block}, which takes into account information from both scans, explicitly putting focus on the prior scan. Our architecture inherently utilizes the feature space difference between the current and prior scans as a weighting factor. This combined representation is then used by the decoder to produce the output segmentation. Fig. \ref{fig:pipeline} shows an overview of our pipeline as well as a schematic of the Difference Weighting Block.

\subsection{Difference Weighting Block}

The Difference Weighting Block acts as a mechanism to weigh and incorporate the temporal differences between images, enhancing the ability of the network to leverage prior scan information for improved segmentation. The architecture involves a series of steps designed to effectively merge latent representations from both the current and prior images. It operates on all resolutions of the U-Net skip connections. Initially, the block calculates the difference between the features $x_c$ extracted from the current image and features $x_p$ obtained from the prior image. Subsequently, an Instance Normalization is applied to the calculated difference, generating a \textit{normalized attention map highlighting areas of change}. The attention map is then multiplied by the features initially derived from the current image. This multiplication process serves as a weighting to emphasize or de-emphasize specific features based on the differences between the two images. It effectively modulates the importance of each feature in the context of changes over time. In contrast to classical attention our block offers a distinct advantage through its lightweight design, which enables 3D operation at all resolutions, and the inherent focus on the differences between inputs. Finally, to ensure a smooth integration, a residual connection is introduced adding the modified features back to the original features extracted from the current image. The output of the Difference Weighting Block are the temporally informed features $x'_c$ used to predict the mask of the current image as shown in Eq. \ref{eq:skipdiff}.

\begin{equation}
   x'_c = x_c \times \mathrm{InstNorm}(x_c - x_p) + x_c
   \label{eq:skipdiff}
\end{equation}

\section{Experiments}

\subsection{Datasets}

To comprehensively assess the efficacy of our model across diverse data, we conducted experiments on two separate longitudinal multiple sclerosis segmentation datasets obtained from distinct medical sites, as shown in Table \ref{tab:datasets}. We use a dataset from the University Medical Center Ljubljana and the Laboratory of Imaging Technologies, University of Ljubljana~\cite{ljubljana_dataset_2}. It contains $1 \mathrm{mm}$ isotropic resolution T1-weighted and FLAIR MR brain images from 162 subjects, each with two to four timepoints, acquired with a 3T Siemens Magnetom Trio MR system. Manual annotations of white-matter lesions were performed through automated segmentation~\cite{ljubljana_dataset_1}, followed by manual corrections by three expert raters~\cite{ljubljana_dataset_2}, with collaborative revisions to achieve consensus. A partial version of the dataset is publicly accessible\footnote[1]{\url{https://github.com/muschellij2/open_ms_data}}, the rest is available upon request. Additionally, we utilize the publicly accessible training split\footnote[2]{\url{https://smart-stats-tools.org/lesion-challenge}} from the ISBI 2015 Longitudinal MS Lesion Segmentation Challenge~\cite{isbi2015challenge} for independent model evaluation. It comprises 5 patients, with an average of 4.4 scans per patient, acquired using a 3T MRI scanner from Philips Medical Systems. Manual annotations of white-matter lesions were conducted by two raters. For consistency with the Ljubljana dataset, we exclusively employ the T1-weighted and FLAIR sequences out of the four available sequences (T1w, T2w, PDw, and FLAIR) for evaluation.

\begin{table}[ht]
    \centering
    \caption{Details of the datasets used for training and evaluation. The dataset split is shown in parenthesis (Training, Test). Note that we use the ISBI 2015 dataset solely as an external testset due to the small size.}
    \label{tab:datasets}
    \begin{tabular}{@{\hspace{1ex}}l@{\hspace{1ex}}l@{\hspace{1ex}}c@{\hspace{1ex}}c@{\hspace{1ex}}c@{\hspace{1ex}}c@{\hspace{1ex}}}
        \toprule
        \multirow{2}{*}{Origin} & \multirow{2}{*}{Modalities} & \multicolumn{2}{c}{Patients} & \multicolumn{2}{c}{Number of scans} \\
         &  & \hspace{.5em}Train & Test & \makebox[4.5em][c]{\hspace{2em}Train} & \makebox[4.5em][c]{Test\hspace{1em}} \\
        \midrule
        Ljubljana~\cite{ljubljana_dataset_2}   & T1, FLAIR  & \hspace{.5em}129 & 33 & \hspace{2em}264 & 67\hspace{1em}
        \\
        ISBI 2015~\cite{isbi2015challenge}    & T1, FLAIR     & \hspace{.5em}- & 5 & \hspace{2em}- & 21\hspace{1em}
        \\
        \bottomrule
    \end{tabular}
\end{table}

\subsection{Evaluation Metrics}

In this study, we assess the efficacy of various models using a comprehensive set of metrics, encompassing volumetric, distance-based, and lesion-centered measures. Specifically, we calculate the Dice coefficient for voxel-level accuracy, the 95\% Hausdorff distance to evaluate boundary discrepancies, and the lesion-based F\textsubscript{1} score as a clinically relevant lesion detection metric\cite{metricsreloaded}. To align with previous evaluation guidelines~\cite{ms_lesion_evaluation}, lesions smaller than $3~\mathrm{mm}^3$ are omitted from both the ground truth and predictions prior to metric computation. Results without size filtering are shown in the Appendix. We compute the lesion-based F\textsubscript{1} score as the harmonic mean of separately calculated lesion-based precision and recall. Following best practice guidelines for medical object detection\cite{nndetection,retinaunet}, we set a threshold of $0.1$ for the Intersection over Union of true positive lesions. Crucially, to account for the varying number of scans per patient and to mitigate potential biases, we average all metrics across patients. The patient-wise averaging ensures a fair and balanced evaluation, reflecting the performance in the clinical scenario.

\subsection{Implementation details}

First, we perform affine registration utilizing FSL FLIRT \cite{fsl-flirt} to align follow-up examinations with the initial baseline scan. This ensures a consistent image space across all scans for a given patient. Following the methodology of nnUNet \cite{Isensee2021}, we automatically determine the optimal U-Net architecture, normalization scheme, and patch size. Training is conducted with a batch size of $2$, utilizing patches of dimensions $128\times160\times128$ and z-score normalization. It is important to note that during preprocessing, we crop all images of an individual patient to the largest nonzero bounding box encompassing all timepoints. For optimization, we employ the default nnUNet training scheme, that is, SGD optimizer with momentum $0.99$, an initial learning rate of $0.01$ and polynomial decay.

\section{Results and Discussion}



\begin{table}[!ht]
  \caption{Comparison of state-of-the-art single timepoint and longitudinal methods for MS lesion segmentation on the Ljubljana dataset~\cite{ljubljana_dataset_2}. \textit{Longitudinal nnUNet} is shown as an ablation. Statistical analysis is given in the Appendix.
  \\\textbf{Bold}: Best performance, \underline{Underlined}: best single or multi timepoint baseline.
  }
  \label{tab:results_ljubljana}
  \centering
  \begin{tabular}{clcccccc}
    \toprule
     &  & \multicolumn{3}{c}{5-fold Cross Val} & \multicolumn{3}{c}{Test Set}  \\
      & Method & DSC~$\uparrow$ & HD95~$\downarrow$ & F\textsubscript{1}~$\uparrow$ & DSC~$\uparrow$ & HD95~$\downarrow$ & F\textsubscript{1}~$\uparrow$ \\
    \midrule
    \multirow{2.15}{*}{\shortstack{Single\\Timepoint}} & nnUNet~\cite{Isensee2021} & \underline{73.37} & \underline{4.51} & \underline{69.83} & \underline{74.16} & \underline{4.95} & \underline{71.27} \\
                                                    & SwinUNETR~\cite{swinunetr} & 73.12 & 4.72 & 69.34 & 74.05 & 4.96 & 70.67 \\
    \midrule
    \multirow{6.15}{*}{\shortstack{Longi-\\tudinal}} & FreeSurfer Samseg~(2023)~\cite{free_surfer_longi} & 62.47 & 12.77 & 37.08 & 45.54 & 15.84 & 33.05 \\
                                                & Denner et.al.~(2020)~\cite{denner2020spatiotemporal} & 67.42 & 5.69 & 61.56 & 69.57 & 5.43 & 65.59 \\
                                                & Szeskin et.al.\tablefootnote[3]{Reimplemented in the nnUNet Framework}~(2022)~\cite{simunet} & 73.24 & 4.87 & 68.45 & 73.44 & 5.11 & 69.08 \\
                                                & Wu et.al. (2023)~\cite{coactseg} &  71.61 & 6.16 & 64.42 & 72.53 & 6.57 & 66.17 \\
    \cmidrule(r){2-8}
                                                & Longitudinal nnUNet~\textit{(Ours)}  & 74.37 & 4.12 & 71.32 & 75.29 & \textbf{4.59} & 72.92 \\
                                                & + Diff. Weighting~\textit{(Ours)}  & \textbf{75.05} & \textbf{3.99} & \textbf{72.72} & \textbf{75.61} & 4.61 & \textbf{73.28} \\

    \bottomrule
  \end{tabular}
\end{table}

\captionsetup[subfigure]{labelformat=empty}
\begin{figure}[!ht]
    \centering
    \begin{subfigure}[t]{0.24\textwidth}
        \centering\includegraphics[width=\textwidth]{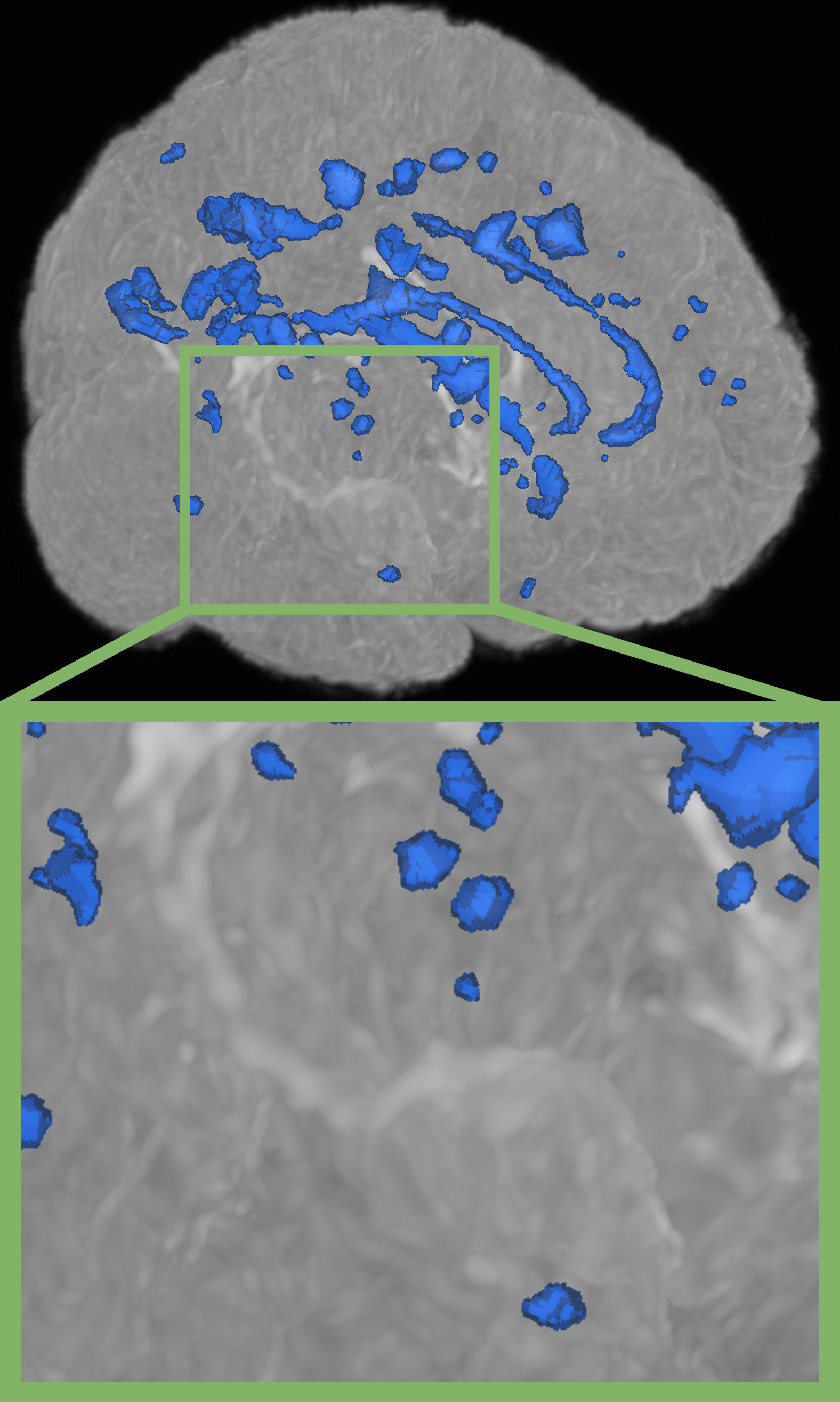}
        \caption{nnUNet\cite{Isensee2021}}
    \end{subfigure}
    \begin{subfigure}[t]{0.24\textwidth}
        \centering\includegraphics[width=\textwidth]{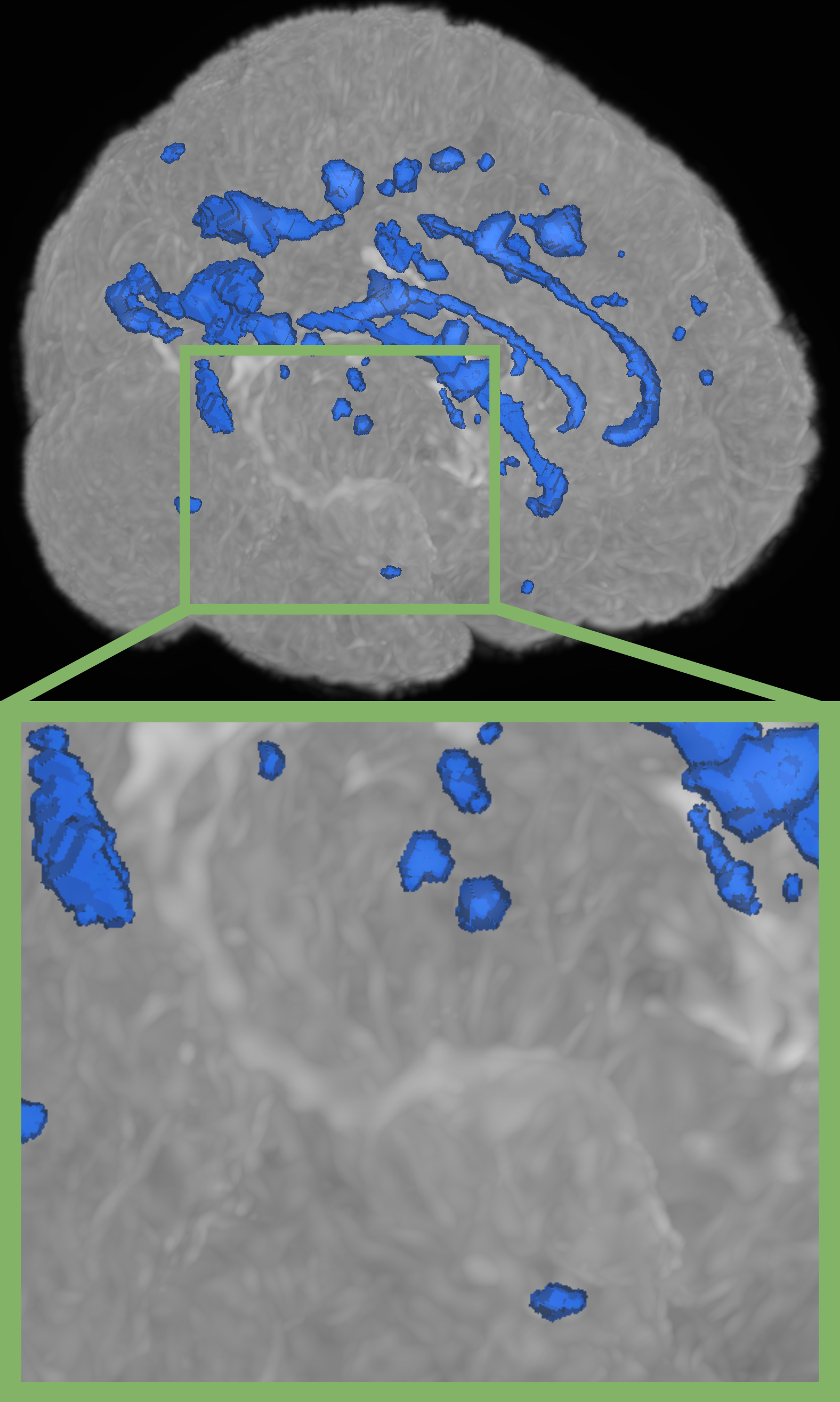}
        \caption{Szeskin et.al.\cite{simunet}}
    \end{subfigure}
    \begin{subfigure}[t]{0.24\textwidth}
        \centering\includegraphics[width=\textwidth]{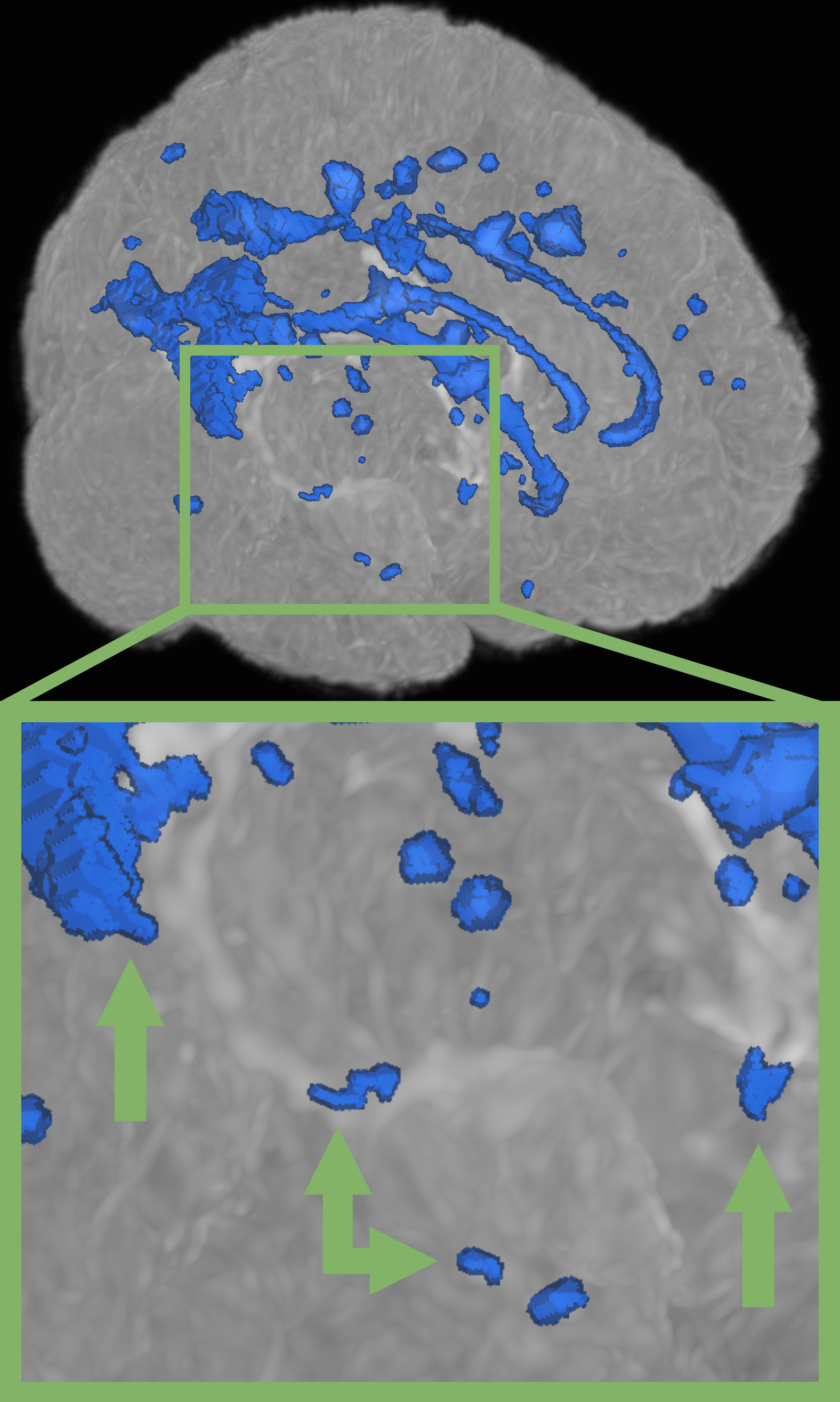}
        \caption{\shortstack{Difference\\Weighting \textit{(Ours)}}}
    \end{subfigure}
    \begin{subfigure}[t]{0.24\textwidth}
        \centering\includegraphics[width=\textwidth]{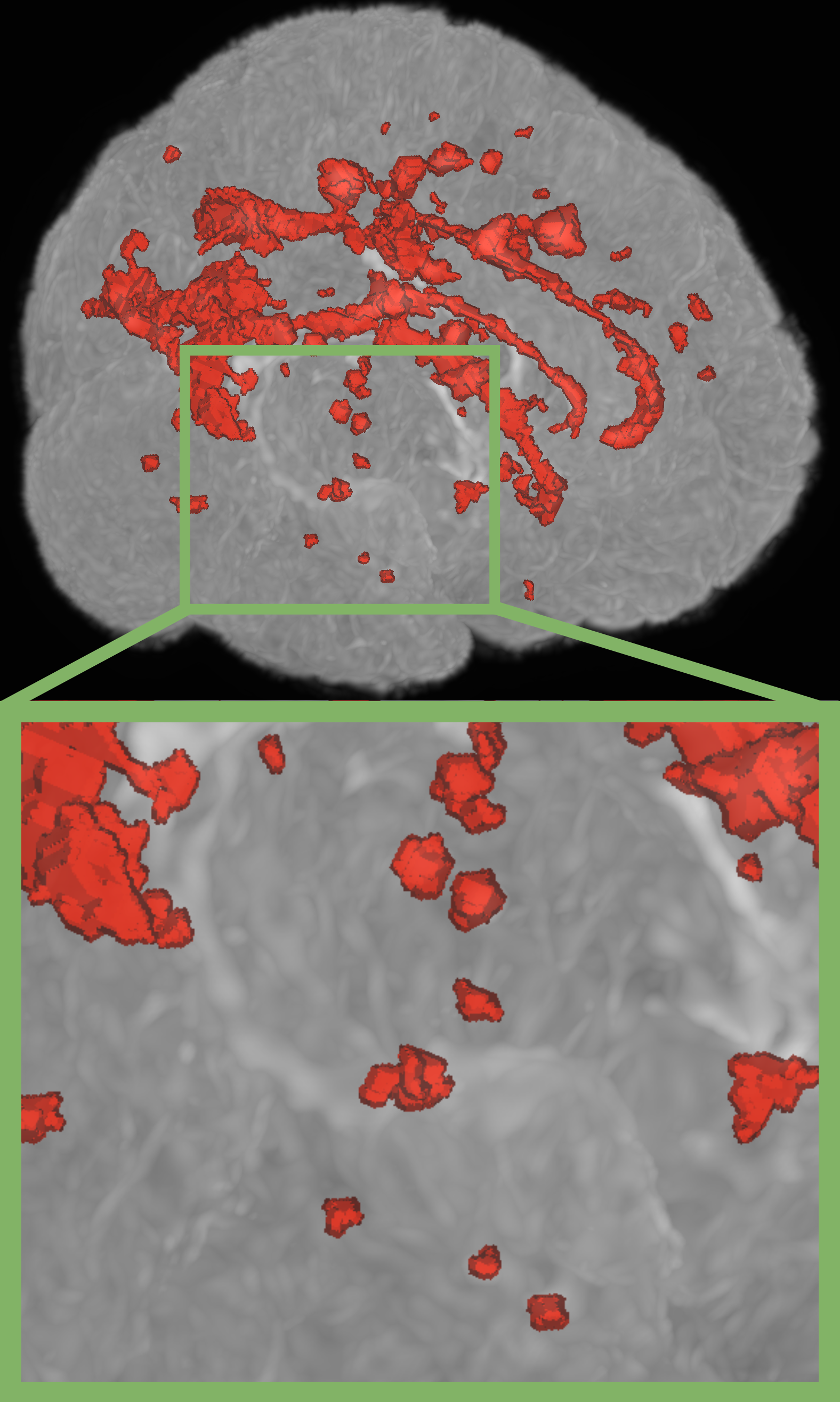}
        \caption{Ground Truth}
    \end{subfigure}
    \caption{Qualitative results on the Ljubljana dataset. Compared to related work, our proposed method demonstrates superior performance in volumetric delineation of MS lesions and successfully identifies lesions missed by other methods.}
    \label{fig:qualitative}
\end{figure}

Table \ref{tab:results_ljubljana} illustrates SOTA performance against several established single and multi timepoint baselines on the Ljubljana dataset. We also show the generalizability of our best models on the ISBI 2015 dataset in Tab. \ref{tab:results_isbi}  respectively. Figure \ref{fig:qualitative} shows additional qualitative results on the Ljubljana dataset. Overall, we demonstrate the following findings:\\

\noindent\textbf{Powerful single timepoint networks outperform existing longitudinal approaches:} nnUNet\cite{Isensee2021}, as the best single timepoint model, achieves a Dice score of 74.16\%, a 95\% Hausdorff Distance of 4.95mm and a lesion-based F\textsubscript{1} score of 71.27\% on the test set of the Ljubljana dataset. The results clearly show that current \textit{longitudinal} approaches consistently struggle to surpass or even reach this performance. The most competitive longitudinal baseline, a reimplementation of Szeskin et.al.'s method \cite{simunet} within the nnUNet framework, still performs worse than nnUNet on all metrics, especially falling short 2.19\% in lesion-based F\textsubscript{1} score. The other two deep learning based baselines from Denner et al.\cite{denner2020spatiotemporal} and Wu el.al.\cite{coactseg} demonstrate even lower performances with large decreases in all metrics. The non deep learning based FreeSurfer Samseg tool\cite{free_surfer_longi} achieves the worst results with a Dice score of 45.54\%, a 95\% Hausdorff distance of 15.84mm and a lesion-based F\textsubscript{1} score of 33.05\%.\\

\noindent\textbf{Pairing powerful single timepoint methods with longitudinal data enables efficient use of additional information:} While nnUNet on single timepoints outperforms inefficient longitudinal methods, the ablation of naively integrating longitudinal data into nnUNet through color channel concatenation (\textit{Longitudinal nnUNet}) delivers enhancements across all investigated metrics when compared to previous longitudinal methods and SOTA single timepoint approaches. This underlines the potential gain longitudinal data can have even just as an additional channel input of a powerful single timepoint network.\\
 
\noindent\textbf{Employing inductive bias via feature differences achieves SOTA performance:} The integration of our novel Difference Weighting method further elevates the model's performance, demonstrating a new state-of-the-art performance. Our proposed method improves upon nnUNet, as the best performing state-of-the-art model, by 1.45\% in Dice score and 0.34mm in 95\% Hausdorff distance. Notably, we see an increase of  2.01\% in lesion-based F\textsubscript{1} score, as the most important clinical metric, indicating that our approach better detects individual lesions. We also surpass the Longitudinal nnUNet indicating that the imposed inductive bias is superior over plain channel-wise concatenation.\\

\noindent\textbf{Benefits from longitudinal data are transferable across datasets:} The evaluation on the ISBI 2015 dataset validates the robustness of our method across datasets as shown in Tab. \ref{tab:results_isbi}. It is worth noting that this dataset, while used for generalization testing, is too small to properly train longitudinal methods. Although slightly underperforming in the 95\% Hausdorff distance, our Difference Weighting approach delivers the best performance in Dice score and lesion-based F\textsubscript{1} score outperforming nnUNet by 1.09\% and 2.71\% respectively. Surpassing FreeSurfer Samseg it thus demonstrates superior overall performance on the ISBI 2015 dataset which illustrates the effective translation of our findings across multiple MS datasets. This highlights the substantial potential of leveraging longitudinal data for improved disease monitoring and assessment.

\begin{table}[!htbp]
  \caption{\textbf{Generalizability to data from other sites.} Independent evaluation of models trained on the Ljubljana dataset~\cite{ljubljana_dataset_2} and the non-deeplearning tool FreeSurfer\cite{free_surfer_longi} on the official ISBI 2015 challenge training dataset.}
  \label{tab:results_isbi}
  \centering
  \begin{tabular}{lcccccc}
    \toprule
    Method & DSC~$\uparrow$  & HD95~$\downarrow$ & F\textsubscript{1}~$\uparrow$ \\
    \midrule
    FreeSurfer Samseg~(2023)~\cite{free_surfer_longi} & 68.50 & 12.42 & 51.97 \\
    nnUNet~\cite{Isensee2021} & 69.74 & \textbf{9.95} & 59.74 \\
    Longitudinal nnUNet~\textit{(Ours)} & 70.77 & 12.13 & 60.64 \\
    + Diff. Weighting~\textit{(Ours)} & \textbf{70.83} & 10.83 & \textbf{62.45}\\
    \bottomrule
  \end{tabular}
\end{table}

\section{Conclusion}

In this paper, we have presented a novel method for multiple sclerosis lesion segmentation explicitly incorporating temporal information using our Difference Weighting Block. This component fuses baseline and follow-up scan features, emphasizing changes for improved segmentation. Experimental results demonstrated our model's superiority over state-of-the-art single timepoint baselines and existing longitudinal methods. Our findings emphasize that leveraging representational differences outperforms naive channel-wise concatenation, with improvements also transferring to unseen datasets. Further research will focus on utilizing this longitudinal approach in addressing additional diseases.

\begin{credits}
\subsubsection{\ackname} The present contribution is supported by the Helmholtz Association under the joint research school "HIDSS4Health – Helmholtz Information and Data Science School for Health". This work was partly funded by Helmholtz Imaging (HI), a platform of the Helmholtz Incubator on Information and Data Science. PV is funded through an Else Kröner Clinician Scientist Endowed Professorship by the Else Kröner Fresenius Foundation (reference number: 2022\_EKCS.17).

\subsubsection{\discintname}
The authors have no competing interests to declare that are relevant to the content of this article.
\end{credits}

\bibliographystyle{splncs04}
\bibliography{references}

\newpage
\appendix

\section{Statistical Analysis}
\label{sec:stat_analysis}

\captionsetup[subfigure]{labelformat=empty}
\begin{figure}[!ht]
    \centering
    \begin{subfigure}[t]{\textwidth}
        \centering\includegraphics[width=\textwidth]{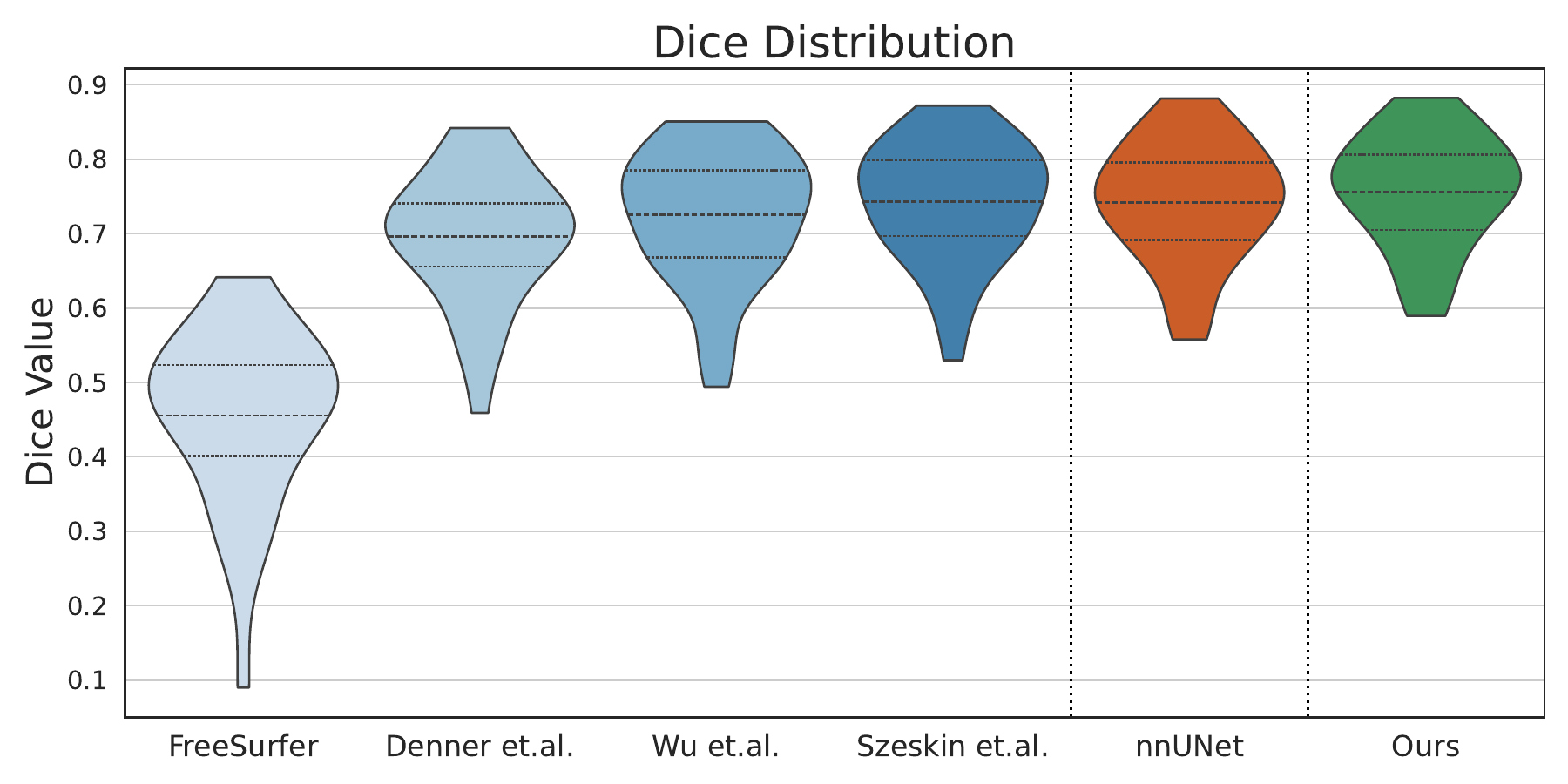}
    \end{subfigure}
    
    \begin{subfigure}[t]{\textwidth}
        \centering\includegraphics[width=\textwidth]{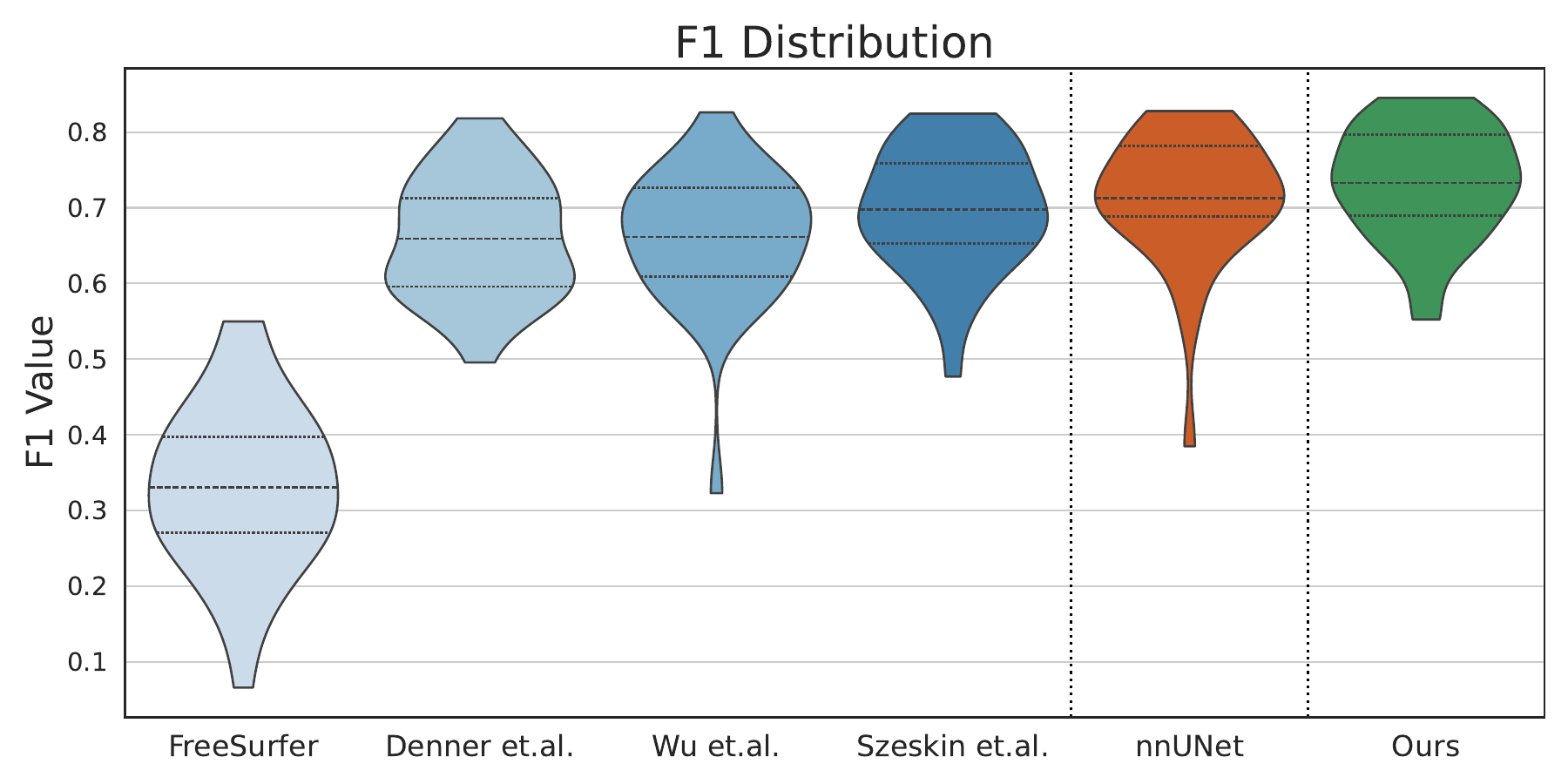}
    \end{subfigure}
    \caption{Distribution of Dice and lesion-based F\textsubscript{1} scores on the Ljubljana test set. The longitudinal baselines are depicted in \textit{blue}, nnUNet as the best single timepoint baseline in \textit{orange}, and our proposed Difference Weighting approach in \textit{green}. The figure depicts the superiority of our proposed method in terms of both the mean and the distribution, particularly evident in the F\textsubscript{1} score. The narrower tails of the distribution indicate more precise segmentations, especially for challenging cases.}
\end{figure}

\newpage
\section{Results without Size Filtering}
\label{sec:no_threshold}

\begin{table}[!ht]
  \caption{Comparison of state-of-the-art single timepoint and longitudinal methods for MS lesion segmentation on the Ljubljana test set \textit{without} size threshold filtering of lesions smaller than $3~\mathrm{mm}^3$.}
  \label{tab:results_ljubljana_no_size}
  \centering
  \begin{tabular}{clccc}
    \toprule
      & Method & DSC~$\uparrow$ & HD95~$\downarrow$ & F\textsubscript{1}~$\uparrow$ \\
    \midrule
    \multirow{2.15}{*}{\shortstack{Single\\Timepoint}} & nnUNet~\cite{Isensee2021} & 74.04 & 4.68 & 67.63 \\
                                                    & SwinUNETR~\cite{swinunetr} & 73.86 & 4.55 & 66.41 \\
    \midrule
    \multirow{6.15}{*}{\shortstack{Longi-\\tudinal}} & FreeSurfer Samseg~(2023)~\cite{free_surfer_longi} & 45.57 & 15.61 & 27.39 \\
                                                & Denner et.al.~(2020)~\cite{denner2020spatiotemporal} & 69.40 & 4.57 & 59.16 \\
                                                & Szeskin et.al.~(2022)~\cite{simunet} & 73.24 & 4.77 & 65.46 \\
                                                & Wu et.al. (2023)~\cite{coactseg} & 72.34 & 6.32 & 59.64 \\
    \cmidrule(r){2-5}
                                                & Longitudinal nnUNet~\textit{(Ours)} & 75.13 & \textbf{4.19} & 68.38 \\
                                                & + Diff. Weighting~\textit{(Ours)} & \textbf{75.41} & 4.29 & \textbf{69.71} \\

    \bottomrule
  \end{tabular}
\end{table}

\end{document}